\documentclass[a4paper,english,prb,twocolumn]{revtex4}
\usepackage[latin1]{inputenc}
\usepackage{graphicx}
\usepackage{epstopdf}
\usepackage{color}
\newcommand{\JQ}[1]{#1}

\makeatletter

\providecommand{\tabularnewline}{\\}
\usepackage{babel}
\makeatother
\begin{document}

\title{Relativistic analysis of the pairing symmetry of the noncentrosymmetric
superconductor LaNiC\emph{$_{2}$}}

\author{Jorge Quintanilla$^{1,2}$}
\author{Adrian D. Hillier$^{2}$}
\author{James F. Annett$^{3}$}
\author{R. Cywinski$^{4}$}

\address{$^{1}$School of Physical Sciences, University of Kent, Canterbury CT2 7NH, United Kingdom}
\address{$^{2}$ISIS Facility, STFC Rutherford Appleton Laboratory, Harwell Science and Innovation Campus, Didcot OX11 0QX, United Kingdom}
\address{$^{3}$H. H. Wills Physics Laboratory, University of Bristol, Tyndall Avenue, Bristol BS8 1TL, United Kingdom}
\address{$^{4}$University of Huddersfield, School of Applied Sciences, Huddersfield HD1 3DH, United Kingdom}

\begin{abstract}
We present a relativistic symmetry analysis of the allowed pairing states in the noncentroymmetric superconductor LaNiC$_2$. The case of zero spin-orbit coupling (SOC) is discussed first and then the evolution of the symmetry-allowed superconducting instabilities as SOC is adiabatically turned on is described. In addition to mixing singlet with triplet pairing, SOC splits some triplet pairing states with degenerate order parameter spaces into non-degenerate pairing states with different critical temperatures. We address the breaking of time-reversal symmetry (TRS) detected in recent muon spin relaxation experiments and show that it is only compatible with such non-unitary triplet pairing states. In particular, an alternative scenario featuring conventional singlet pairing with a small admixture of triplet pairing is shown to be incompatible with the experimental data.
\end{abstract}

\maketitle

\section{Introduction}

%Since the discovery of superconductivity in the heavy fermion compound
%CePt3Si \cite{bauer_heavy_2004} \textbf{{[}...] {[}}\textbf{\underbar{James,}}
%\textbf{could you please write something generic about noncentrosymmetric
%Superocnductors? Anything I write hsere would be just too similar to
%the first paragraphs of Ref.~\onlinecite{hillier_evidence_2009}] }

Noncentrosymmetric superconductors have been a subject of considerable
interest since the discovery of superconductivity in the heavy fermion
material CePt$_3$Si.\cite{bauer_heavy_2004} In particular it is the unique
property of such noncentrosymmetric superconductors that in the presence
of spin-orbit coupling (SOC) both spin singlet and spin triplet Cooper
pairs can, and  must, coexist within a single material. \JQ{This is quite general as, while in the complete absence of SOC the two kinds
of pairing are distinguished by their different behavior under rotations
in spin space,} once SOC is finite then spin and space rotations
cannot be separated, and it is only the parity of the Cooper pair wave
function under spatial inversion, P, which separates spin singlet (even)
from spin triplet (odd) states.\cite{anderson_1984,volovik_1984,ueda_rice_1985,ozaki_machida_omi_1985,blount_1985}
In  a noncentrosymmetric superconductor there is no lattice center of
inversion, and so the parity operator, P, is not a well defined symmetry
of the crystal, leading to mixing of singlet and triplet pairing states
within a single material. An interesting analogy can be made with particle
physics where the mixing of neutrino flavors is induced by violation of CP
symmetry.\cite{Weinberg} The implication is that in noncentrosymmetric 
superconductors the order parameter is always unconventional. On the other hand the experimental situation is quite complex as some noncentrosymmetric superconductors such as CePt$_3$Si are additionally strongly correlated while the superconducting state of others such as Li$_2$Pd$_3$B, BaPtSi$_3$ \JQ{or Re$_3$W} appears to feature pure singlet pairing.\JQ{\cite{Yuan_Li2Pd3B_2006,Bauer_BaPtSi3_2009,Zuev_Re3W_2007}} 

An important recent development has been the observation, through
zero-field muon spin resonance ($\mu$SR), of time reversal symmetry (TRS)
breaking at the superconducting instability of
LaNiC$_{2}$.\cite{hillier_evidence_2009}   Superconductivity in this
intermetallic compound\cite{bodak_marusin} was discovered in the mid nineties with critical
temperature $T_c=2.7{\rm K}$.\cite{lee_superconductivity_1996} There was
some discussion of whether it was a type II or a dirty type I
superconductor\cite{pecharsky_low-temperature_1998} and the possibility
that the symmetry of the superconducting order parameter was
unconventional was
debated.\cite{lee_superconductivity_1996,pecharsky_low-temperature_1998}
At the time, however, the lack of inversion symmetry was largely
overlooked. In contrast, very recently there has been a surge of
experimental
\cite{hillier_evidence_2009,sung_antipressure_2008,t.f._liao_alloying_2009}
and theoretical
\cite{hase_electronic_2009,hillier_evidence_2009,laverock_electronic_2009,subedi_electron-phonon_2009}
work on this system. Some of this has been motivated by the results in
Ref.~\onlinecite{hillier_evidence_2009} which constitute very strong and
direct evidence of unconventional pairing. In addition to this dramatic dependences of $T_c$ \JQ{on Cu, Y and Th} doping have been identified.\JQ{\cite{sung_antipressure_2008,t.f._liao_alloying_2009,Lee1997323,Lee1997433}} 

Two broad and mutually-exclusive scenarios have been proposed to describe the breaking of TRS in the superconducting state of LaNiC$_2$.\cite{hillier_evidence_2009,subedi_electron-phonon_2009} In the first scenario, which is based on group-theoretical considerations,\cite{hillier_evidence_2009} the superconducting order parameter is intrinsically unconventional: a non-unitary triplet pairing state. In the second scenario, based on first principles calculations,\cite{subedi_electron-phonon_2009} LaNiC$_2$ is essentially a conventional superconductor but a small amount of triplet pairing is induced by SOC, as described above, and is responsible for the observed breaking of TRS. 
Unfortunately both the group theoretical analysis of Ref.~\onlinecite{hillier_evidence_2009} and the first principles calculations of Ref.~\onlinecite{subedi_electron-phonon_2009} ignore relativistic effects. 
It is therefore unclear whether any of the eight superconducting instabilities that are allowed by symmetry and that preserve TRS\cite{hillier_evidence_2009} acquire a TRS breaking component when SOC is adiabatically turned on. More specifically it is not known whether the conventional superconducting state assumed in Ref.~\onlinecite{subedi_electron-phonon_2009}, which does not break TRS, can acquire the necessary TRS breaking component in this way.  
Indeed it is well known\cite{mineev_introduction_1999} that TRS breaking requires a superconducting order parameter with degeneracy. However no such degeneracy should occur in an orthorhombic crystal with finite SOC. In the present work we address this question directly by extending the previous symmetry analysis\cite{hillier_evidence_2009} to include the effect of SOC. The more general analysis that we present here allows us to conclude that the observation of time-reversal symmetry breaking at $T_c$ is not compatible with a conventional mechanism of the type proposed in Ref.~\onlinecite{subedi_electron-phonon_2009}.

To address the pairing symmetry in the LaNiC$_2$ crystal structure\cite{bodak_marusin} we first consider, following the original analysis,\cite{hillier_evidence_2009} the possible pairing states if spin-orbit interaction is negligible. We then study how these states evolve when perturbed by SOC. In particular we note in this paper that simply a combination of $s$-wave pairing and noncentrosymmetric crystal structure does not automatically lead to time-reversal symmetry breaking at $T_c$. It turns out that the low symmetry of the orthorhombic $Amm2$ structure leads to only a small number of time-reversal symmetry breaking states in the absence of SOC, all of which have degeneracy which is lifted when spin-orbit interaction is finite. Therefore the observation of time-reversal symmetry breaking at $T_c$ provides a very strong constraint on the pairing state and is not naturally consistent with the conventional electron-phonon pairing mechanism or $s$-wave pairing. Instead, the observation is only compatible with SOC being small and with the system entering a non-unitary triplet pairing state at $T_c$. 

%In this paper we generalize the symmetry analysis presented
%in Ref.~\onlinecite{hillier_evidence_2009} to the case where spin-orbit
%coupling (SOC) cannot be neglected. 
Our arguments are based on group
theory and in that spirit the present analysis of the pairing symmetry
in LaNiC$_{2}$ does not rely on any specific assumptions about the
origin of the pairing interaction, the band structure or the strength
of SOC\textbf{.} The method is very well-established and has been
very successful in the past for many other superconductors with a centre of inversion,
e.g. the cuprates.\cite{james_f_annett_symmetry_1990} More recently similar methods have been applied to
noncentrosymmetric superconductors.\cite{2004-Sergienko,samokhin_cept3si:unconventional_2004,samokhin_erratum:_2004} We will
nevertheless describe some of the main arguments in considerable detail to highlight the issue of TRS breaking, both in the presence and absence of SOC, as well as the features specific to the point symmetry of LaNiC$_2$.

\section{Symmetry analysis in the absence of spin-orbit coupling\label{sec:weak_SOC}}

The possible symmetries of the superconducting instability in LaNiC$_{2}$
assuming that SOC can be neglected were enumerated in Ref.~\onlinecite{hillier_evidence_2009}.
In this section we give the details of the derivation emphasizing
the similarities and differences with the case where there is a centre
of inversion. In the absence of spin-orbit coupling, the point group
$G$ is \begin{equation}
G=G_{c}\times SO\left(3\right)\label{eq:GcxSO3}\end{equation}
where $\times$ represents the direct product, $G_{c}$ is the point
group of the crystal structure and $SO\left(3\right)$ represents
all spin rotations. The irreducible representations therefore have
the form $\Gamma=\Gamma^{c}\times\Gamma^{s}$ where $\Gamma^{c}$
and $\Gamma^{s}$ are irreducible representations of $G_{c}$ and
$SO\left(3\right)$, respectively (in principle, the full
space group of the crystal must be taken into account; however we
assume that the translational symmetries are the same above and below
$T_{c}$, so it is enough to refer to the point group).
A basis of $\Gamma$ is given by
the functions $\hat{\Gamma}_{mn}\left(\mathbf{k}\right)=\Gamma_{m}^{c}\left(\mathbf{k}\right)\hat{\Gamma}_{n}^{s}$
where $\left\{ \Gamma_{m}^{c}\left(\mathbf{k}\right)\right\} _{m=1,\ldots,d_{\Gamma^{c}}}$
forms a basis of $\Gamma^{c}$ and $\left\{ \hat{\Gamma}_{n}^{s}\right\} _{n=1,\ldots,d_{\Gamma^{s}}}$
forms a basis of $\Gamma^{s}$. The dimensionality of $\Gamma$ is
$d_{\Gamma}=d_{\Gamma^{c}}d_{\Gamma^{s}}$. The gap function just
below $T_{c}$ is thus 
$
%\begin{equation}
\hat{\Delta}\left(\mathbf{k}\right)=\sum_{m=1}^{d_{\Gamma^{c}}}\sum_{n=1}^{d_{\Gamma^{s}}}\eta_{m,n}\Gamma_{m}^{c}\left(\mathbf{k}\right)\hat{\Gamma}_{n}^{s}.\label{eq:Dexp-noSOC}
%\end{equation}
$

The spin rotation group $SO\left(3\right)$ is the same for all crystals.
As is well known it has two irreducible representations (irreps). The first
of these is the singlet representation, of dimension 1. This corresponds
to order parameters of the form $\hat{\Delta}\left(\mathbf{k}\right)=\sum_{m=1}^{d_{\Gamma^{c}}}\eta_{m,0}\Gamma_{m}^{c}\left(\mathbf{k}\right)\hat{\Gamma}_{\mbox{singlet}}^{s}.$
Crucially, $\hat{\Gamma}_{\mbox{singlet}}^{s}=-\left(\hat{\Gamma}_{\mbox{singlet}}^{s}\right)^{T}$
meaning that we must have $\Gamma_{m}^{c}\left(\mathbf{k}\right)=\Gamma_{m}^{c}\left(-\mathbf{k}\right)$.
Thus for singlet order parameters only the first term in 
\begin{equation}
\hat{\Delta}\left(\mathbf{k}\right)=\Delta\left(\mathbf{k}\right)i\hat{\sigma}_{y}+\left[\mathbf{d}\left(\mathbf{k}\right).\left(\hat{\sigma}_{x},\hat{\sigma}_{y},\hat{\sigma}_{z}\right)\right]\hat{\sigma}_{y},
\label{eq:gapf2}
\end{equation}
is finite.
%:
%\begin{equation}
%\hat{\Delta}\left(\mathbf{k}\right)=\Delta_{0}\left(\mathbf{k}\right)i\hat{\sigma}_{y}.
%\label{eq:singlet}
%\end{equation}
 The second irrep of $SO\left(3\right)$ is the triplet representation,
of dimension 3. For it we thus have $\hat{\Delta}\left(\mathbf{k}\right)=\sum_{m=1}^{d_{\Gamma^{c}}}\sum_{n=-1,0,+1}\eta_{m,n}\Gamma_{m}^{c}\left(\mathbf{k}\right)\hat{\Gamma}_{\mbox{triplet},n}^{s}.$
Moreover we have $\hat{\Gamma}_{\mbox{triplet},n}^{s}=\left(\hat{\Gamma}_{\mbox{triplet},n}^{s}\right)^{T}$
whereby the $G_{c}$ basis functions must be odd, $\Gamma_{m}^{c}\left(\mathbf{k}\right)=-\Gamma_{m}^{c}\left(-\mathbf{k}\right)$,
meaning that for triplet pairing (\ref{eq:gapf2}) has only the second
term.
%:\begin{equation}
%\hat{\Delta}\left(\mathbf{k}\right)=\left[\mathbf{d}\left(\mathbf{k}\right).\left(\hat{\sigma}_{x},\hat{\sigma}_{y},\hat{\sigma}_{z}\right)\right]\hat{\sigma}_{y}.\label{eq:triplet}\end{equation}

The above results are very well known from the group theory analysis
of centrosymmetric superconductors.\cite{james_f_annett_symmetry_1990,sigrist_phenomenological_1991,mineev_introduction_1999}
They are also valid in the noncentrosymmetric case as long \JQ{as} SOC can
be neglected. In particular, the pairing symmetry must be purely of
the singlet 
%(\ref{eq:singlet}) 
or triplet 
%(\ref{eq:triplet}) 
type
in the limit in which SOC does not play a role. The only difference
with the case of centrosymmetric superconductors is that in a noncentrosymmetric
superconductor the irreps of the crystal point group do not have distinct
symmetries under inversion, so each of them is compatible with both
singlet and triplet pairing. Thus in LaNiC$_{2}$, where $G_{c}=C_{2v}$,
each of the four irreps $A_{1},A_{2},B_{1}$ and $B_{2}$ [Table I in Ref.~\onlinecite{hillier_evidence_2009}]
is compatible with singlet and triplet superconducting instabilities.
Since in this case all four irreps of $G_{c}$ are one-dimensional,
this leads to a total of 12 possible instabilities: 4 in the singlet
channel and 8 in the triplet channel (see Ref.~\onlinecite{hillier_evidence_2009} for details). The possible
symmetries of the gap function are reproduced in Table \ref{tab:C2v_noSOC} here
for completeness. Note that the non-unitary triplet pairing instabilities
$^{3}A_{1}\left(b\right),{}^{3}A_{2}\left(b\right),{}^{3}B_{1}\left(b\right),{}^{3}B_{2}\left(b\right)$
are the only ones that break TRS, leading to the conclusion
that the superconducting state just below $T_{c}$ features nonunitary
triplet pairing.\cite{hillier_evidence_2009} As noted in that reference
one of these four forms of the gap function has the same point group
symmetry as the crystal, which would not have been possible for triplet
pairing in a centrosymmetric superconductor. The other three break
additional symmetries. In the following section we analyse how this
conclusion is affected by the inclusion of SOC in the analysis. 

\begin{table}
\begin{tabular}{|c|c|c|}
\hline 
Irrep of $SO\left(3\right)\times C_{2v}$&
$\Delta_{0}\left(\mathbf{k}\right)$&
$\mathbf{d}\left(\mathbf{k}\right)$\tabularnewline
\hline
\hline 
$^{1}A_{1}$&
$1$&
0\tabularnewline
\hline 
$^{1}A_{2}$&
$XY$&
0\tabularnewline
\hline 
$^{1}B_{1}$&
$XZ$&
0\tabularnewline
\hline 
$^{1}B_{2}$&
$YZ$&
0\tabularnewline
\hline 
$^{3}A_{1}(a)$&
0&
$\left(0,0,1\right)Z$\tabularnewline
\hline 
$^{3}A_{2}(a)$&
0&
$\left(0,0,1\right)XYZ$\tabularnewline
\hline 
$^{3}B_{1}(a)$&
0&
$\left(0,0,1\right)X$\tabularnewline
\hline 
$^{3}B_{2}(a)$&
0&
$\left(0,0,1\right)Y$\tabularnewline
\hline 
$^{3}A_{1}(b)$&
0&
$\left(1,i,0\right)Z$\tabularnewline
\hline 
$^{3}A_{2}(b)$&
0&
$\left(1,i,0\right)XYZ$\tabularnewline
\hline 
$^{3}B_{1}(b)$&
0&
$\left(1,i,0\right)X$\tabularnewline
\hline 
$^{3}B_{2}(b)$&
0&
$\left(1,i,0\right)Y$\tabularnewline
\hline
\end{tabular}

\caption{\label{tab:C2v_noSOC}Possible symmetries of the gap function of
LaNiC$_{2}$ just below $T_{c}$ in the case where SOC can be neglected,
written in terms of $\Delta_{0}\left(\mathbf{k}\right)$ and $\mathbf{d}\left(\mathbf{k}\right)$
in Eq.~(\ref{eq:gapf2}). Each of the functions $X,Y,Z$ depend on
the wave vector $\mathbf{k}$ and they have the same symmetries under
the opretaions of the point group $C_{2v}$ as its three components
$k_{x},k_{y}$ and $k_{z}$, respectively.}
\end{table}

\section{Symmetry analysis in the presence of spin-orbit coupling\label{sec:strong_SOC}}

Now suppose that spin-orbit coupling is strong enough that it cannot
be neglected. Then, as in the case of centrosymmetric superconductors,
$G=G_{c,J},$ which is the ``double group'' obtained by appending
to each rotation carried out on the coordinates in $G_{c}$ an equivalent
operation carried out on the spins. Take, for example, the reflection
through the $x-z$ plane contained in the point group of the LaNiC$_{2}$
crystal structure, $C_{2v}$. This is $\sigma_{v}=IC_{2}^{y}$ where
$I$ represents inversion through the central point and $C_{2}^{y}$
a rotation by $180^{\mbox{o}}$ around the $y$ axis. Then $G_{c,J}$
contains the similar operation, $\sigma_{v,J}$, involving this reflection
as well as a $C_{2}^{y}$ rotation carried out on the spins (i.e. a rotation of the $\bf{d}$ vector). 
%Thus
%the operation of $\sigma_{v,J}$ on a general gap function is
%\begin{widetext}
%\begin{eqnarray}
%\sigma_{v,J}\hat{\Delta}\left(k_{x},k_{y},k_{z}\right) & = & %\Delta_{0}\left(k_{x},-k_{y},k_{z}\right)i\hat{\sigma}_{y}+I\left[\left(-d_{x}\left(-k_{x},k_{y},-k_{z}\right),d_{y}\left(-k_{x},k_{y},-k_{z}\right),-d_{z}\left(-k_{x},k_{y},-k_{z}\right)\right).\left(\hat{\sigma}_{x},\hat{\sigma}_{y},\hat{\sigma}_{z}\right)\right]\hat{\sigma}_{y}\nonumber %\\
% & = & %\Delta_{0}\left(k_{x},-k_{y},k_{z}\right)i\hat{\sigma}_{y}+\left[\left(-d_{x}\left(k_{x},-k_{y},k_{z}\right),d_{y}\left(k_{x},-k_{y},k_{z}\right),-d_{z}\left(k_{x},-k_{y},k_{z}\right)\right).\left(\hat{\sigma}_{x},\hat{\sigma}_{y},\hat{\sigma}_{z}\right)\right]\hat{\sigma}_{y}.\label{eq:sigma_v}\end{eqnarray}
%\end{widetext}
%Eq.~(\ref{eq:Dexp-noSOC}) is no longer valid. 
%Instead,
The gap function just below $T_{c}$ is now 
%a linear combination of a
%set of basis functions of one of the irreducible representations of
%$G_{c,J}$,
%\begin{equation}
$
\hat{\Delta}\left(\mathbf{k}\right)=\sum_{i=1}^{d_{\Gamma}}\eta_{i}\hat{\Gamma}_{i}\left(\mathbf{k}\right)$,
%\label{eq:Dexp}\end{equation}
 where $\hat{\Gamma}_{i}\left(\mathbf{k}\right)$
is the $i^{\mbox{th}}$ basis function of the irrep $\Gamma$ of $G_{c,J}$.
In general, unlike the case of vanishing SOC, the gap function is
not of the singlet or triplet forms. 
%(\ref{eq:singlet},\ref{eq:triplet}).
Note, however, that such mixture of the singlet and triplet channels
occurs only when \emph{both} of the following conditions are met:
(i) there is no centre of inversion \emph{and} (ii) SOC cannot be
neglected. As has been extensively remarked \cite{gorkov_superconducting_2001,frigeri_erratum:_2004,frigeri_superconductivity_2004,bauer_heavy_2004}
this makes noncentrosymmetric superconductors special in that SOC
has a more dramatic effect on the pairing symmetry than it has in
centrosymmetric superconductors.\cite{james_f_annett_symmetry_1990,sigrist_phenomenological_1991}
On the other hand that is quite different from saying that SOC
has to be strong in these systems. Indeed, as we will see shortly
in the case of LaNiC$_{2}$ it is difficult to reconcile the observation
of TRS breaking \cite{hillier_evidence_2009} with SOC being strong. 

Through SOC, spin rotations cease to be independent degrees of freedom.
Thus unlike the case of zero SOC the irreps of $G_{c,J}$ are in one-to-one
correspondence to those of $G_{c}$. For LaNiC$_{2}$ this leads to
a dramatic reduction in the number of symmetry-allowed superconducting
instabilities of the normal state from 12 when SOC can be neglected
(see above) to only 4, corresponding to the 4 irreps of the point
group of the crystal structure. The basis functions depend both on
$\mathbf{k}$ \JQ{and the} spin indices (i.e. they are matrices), just
like the basis functions of the irreps of $G_{c}\times SO\left(3\right)$.
Constructing the four symmetry operations $E_J,C_{2,J},\sigma_{v,J}$ and $\sigma_{v,J}'$ in the way described above 
%Using (\ref{eq:sigma_v}) and three similar equations to describe
%the four symmetry operations $E,C_{2},\sigma_{v}$ and $\sigma_{v}'$
one can find a set of basis functions that is compatible with
the group's character table [Table I in Ref.~\onlinecite{hillier_evidence_2009}]. One such set is given in Table \ref{tab:C2v_SOC}. The $A,B,C,D$ coefficients should be determined by a microscopic theory but should be real. Note that, as a direct result of all the irreps of $G_{c}$ being one-dimensional [Table I in Ref.~\onlinecite{hillier_evidence_2009}], all the possible order parameters just below $T_{c}$ are one-dimensional, too. Since a one-dimensional order parameter cannot break TRS \cite{mineev_introduction_1999} we are led to the inescapable conclusion that \emph{the superconducting instability in LaNiC$_2$ can only break
TRS if SOC is negligible.} In view of the experimental observation
of TRS breaking,\cite{hillier_evidence_2009} this suggests that
the effect of SOC on the superconductivity must be small and confirms
our original conclusion,\cite{hillier_evidence_2009} reached on
the basis of a nonrelativsitic analysis, of nonunitary triplet pairing. 

\begin{table}
\begin{tabular}{|c|c|c|}
\hline 
Irrep of $C_{2v,J}$&
$\Delta_{0}\left(\mathbf{k}\right)$&
$\mathbf{d}\left(\mathbf{k}\right)$\tabularnewline
\hline
\hline 
$A_{1}$&
$A$&
$\left(BY,CX,DXYZ\right)$\tabularnewline
\hline 
$A_{2}$&
$AXY$&
$\left(BX,CY,DZ\right)$\tabularnewline
\hline 
$B_{1}$&
$AXZ$&
$\left(BXYZ,CZ,DY\right)$\tabularnewline
\hline 
$B_{2}$&
$AYZ$&
$\left(BZ,CXYZ,DX\right)$\tabularnewline
\hline
\end{tabular}

\caption{\label{tab:C2v_SOC}Possible symmetries of the gap function of LaNiC$_{2}$
just below $T_{c}$ in the case where SOC cannot be neglected. $A,B,C,D$
denote four $\mathbf{k}$-independent quantities with the same phase.
All other notations as in Table \ref{eq:Dexp-noSOC}. }
\end{table}

Note that the case of the orthorhombic symmetry group $C_{2v}$ appropriate for LaNiC$_2$ is
quite different from the tetragonal $C_{4v}$ appropriate to CePt$_3$Si\cite{2004-Sergienko,samokhin_cept3si:unconventional_2004,samokhin_erratum:_2004}. For $C_{4v}$ one of the irreducible representations is two-dimensional, so time-reversal symmetry breaking is allowed even in the presence of strong SOC. The point group studied here is also somewhat different from the monoclinic $C_2$, studied by Sergienko and Curnoe.\cite{2004-Sergienko} In this case there is only one two-fold rotation axis, and hence only two irreducible representations, $A_1$ and $A_2$, both
one-dimensional. Nevertheless the general pattern of possible symmetry breakings for $C_{2}$ is similar to 
those given in Tables I and II. Under $C_2$ the $A_1$ representation is equivalent to both $A_1$ and 
$A_2$ of $C_{2v}$, while the $A_2$ representation of $C_2$ is equivalent to 
$B_1$ and $B_2$ under $C_{2v}$.

\section{Spin-orbit coupling-induced splitting of the superconducting instability}

Our main conclusion so far is that the observation of TRS symmetry
breaking implies that SOC must be very weak, for no TRS breaking superconducting
instability of the normal state is compatible with the crystal's symmetry
in the presence of SOC. One the other hand, a small amount of SOC
must be present in any crystal, which raises the question of how the
results of Secs.~\ref{sec:weak_SOC} and \ref{sec:strong_SOC} can
be reconciled. To clarify this we consider the evolution of the instability
as a small amount of SOC is adiabatically turned on. 

\begin{figure}
\includegraphics[width=1.0\columnwidth,keepaspectratio]{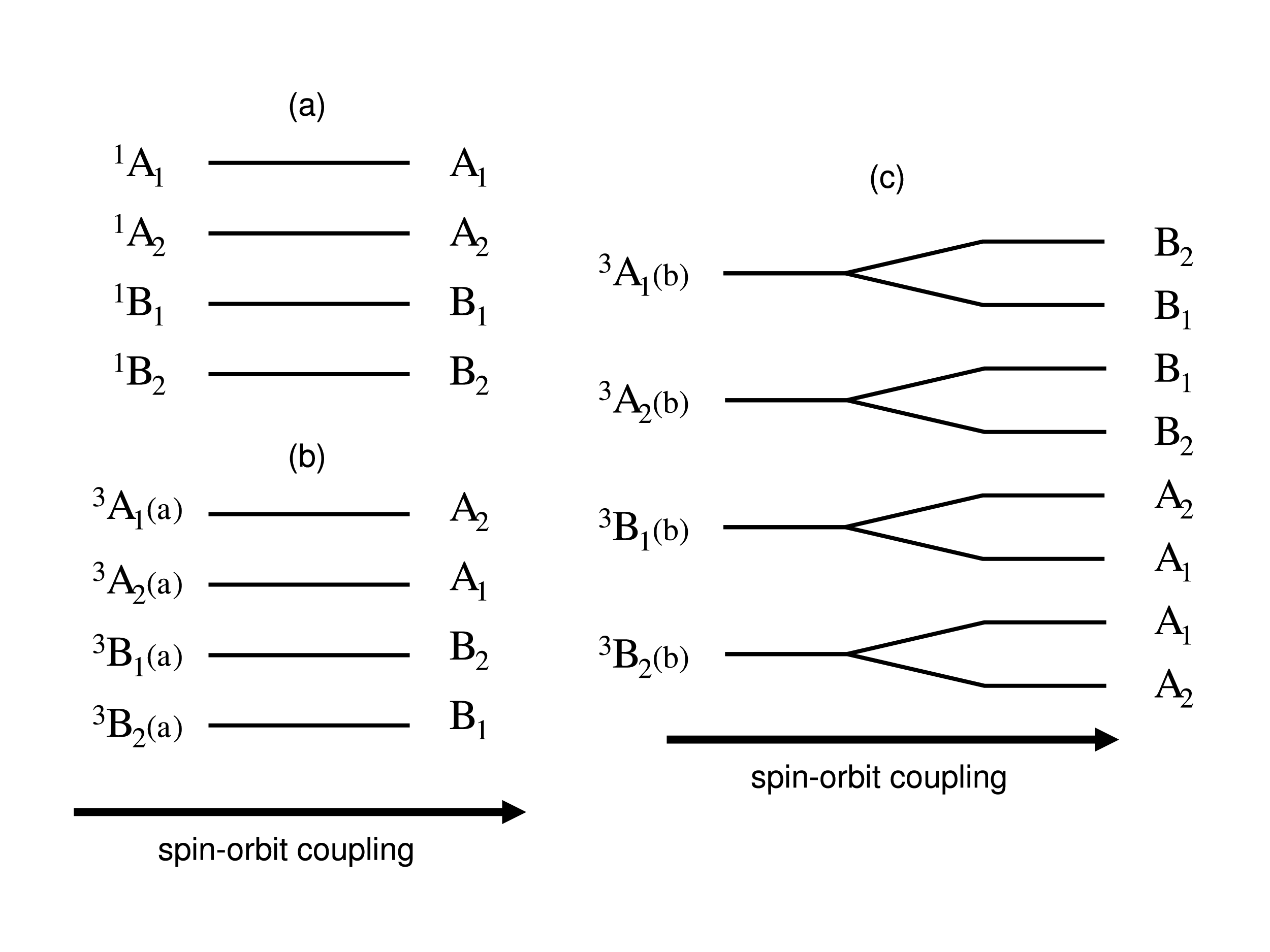}
\caption{\label{fig:Evolution}Evolution of all the superconducting instabilities
of the normal state of LaNiC$_{2}$ allowed by symmetry in the absence
of SOC as the latter is adiabatically turned on: (a) singlet pairing
instabilities; (b) unitary triplet pairing instabilities; (c) non-unitary
triplet pairing instabilities. The relative temperatures of the different
instabilities in this diagram are arbitrary.}
\end{figure}

Each of the symmetry-allowed
superconducting instabilities listed in Table \ref{tab:C2v_noSOC}
will evolve into one of those listed in Table \ref{tab:C2v_SOC}, as shown in Fig.~\ref{fig:Evolution}.
To ascertain the relationships depicted in the figure, we must express the gap function given in
Table \ref{tab:C2v_noSOC} as a linear combination of those in Table
\ref{tab:C2v_SOC}. Such linear combinations are unique. In particular,
the $\mathbf{k}$-dependences of the gap function just below the singlet
superconducting instabilities are given by\begin{eqnarray}
\hat{\Gamma}_{^{1}A_{1}}\left(\mathbf{k}\right) & = & \left.\hat{\Gamma}_{A_{1}}\left(\mathbf{k}\right)\right|_{A,B,C,D=1,0,0,0}\label{eq:adiab_evol_1A1}\\
\hat{\Gamma}_{^{1}A_{2}}\left(\mathbf{k}\right) & = & \left.\hat{\Gamma}_{A_{2}}\left(\mathbf{k}\right)\right|_{A,B,C,D=1,0,0,0}\label{eq:adiab_evol_1A2}\\
\hat{\Gamma}_{^{1}B_{1}}\left(\mathbf{k}\right) & = & \left.\hat{\Gamma}_{B_{1}}\left(\mathbf{k}\right)\right|_{A,B,C,D=1,0,0,0}\label{eq:adiab_evol_1B1}\\
\hat{\Gamma}_{^{1}B_{2}}\left(\mathbf{k}\right) & = & \left.\hat{\Gamma}_{B_{2}}\left(\mathbf{k}\right)\right|_{A,B,C,D=1,0,0,0}\label{eq:adiab_evol_1B2}\end{eqnarray}
Thus the $^{1}A_{1},^{1}A_{2},^{1}B_{1}$ and $^{1}B_{2}$ instabilities
evolve into instabilities with $A_{1},A_{2},B_{1}$ and $B_{2}$ symmetries,
respectively. Moreover adiabatic continuity in the limit of vanishing
SOC places constraints on the coefficients $A,B,C,D$ in Table \ref{tab:C2v_SOC}:
in order that the coefficients $A,B,C$ vanish in the limit of zero
SOC, it is necessary for them to be small, compared to $D$, when
SOC is weak but finite. By this mechanism a small triplet component
{[}i.e. a finite $\mathbf{d}\left(\mathbf{k}\right)$] could be induced
in an otherwise singlet superconductor by the action of SOC alone.
Note, however, that such triplet component does not break TRS. This
is at variance with the claim made in Ref.~\onlinecite{subedi_electron-phonon_2009},
as we discuss in detail in Sec.~\ref{sec:Discussion}.

Similarly, the $\mathbf{k}$-dependences of the gap function just
below the four unitary triplet pairing instabilities are also in one-to-one
correspondence with those of the relativistically-allowed ones:\begin{eqnarray}
\hat{\Gamma}_{^{3}A_{1}(a)}\left(\mathbf{k}\right) & = & \left.\hat{\Gamma}_{A_{2}}\left(\mathbf{k}\right)\right|_{A,B,C,D=0,0,0,1}\label{eq:adiab_evol_3A1a}\\
\hat{\Gamma}_{^{3}A_{2}(a)}\left(\mathbf{k}\right) & = & \left.\hat{\Gamma}_{A_{1}}\left(\mathbf{k}\right)\right|_{A,B,C,D=0,0,0,1}\label{eq:adiab_evol_3A2a}\\
\hat{\Gamma}_{^{3}B_{1}(a)}\left(\mathbf{k}\right) & = & \left.\hat{\Gamma}_{B_{2}}\left(\mathbf{k}\right)\right|_{A,B,C,D=0,0,0,1}\label{eq:adiab_evol_3B1a}\\
\hat{\Gamma}_{^{3}B_{2}(a)}\left(\mathbf{k}\right) & = & \left.\hat{\Gamma}_{B_{1}}\left(\mathbf{k}\right)\right|_{A,B,C,D=0,0,0,1}\label{eq:adiab_evol_3B2a}\end{eqnarray}
Finally, for the four non-unitary triplet pairing instabilities the
situation is somewhat more complicated. Since they break TRS, they cannot evolve smoothly into one of the four
symmetry-allowed instabilities as SOC is turned on, as all of them
preserve TRS. Indeed the gap matrix just below $T_{c}$ is a linear
combination of two of the forms allowed in the presence of SOC:\begin{widetext}
\begin{eqnarray}
\hat{\Gamma}_{^{3}A_{1}(b)}\left(\mathbf{k}\right) & = & \left.\hat{\Gamma}_{B_{2}}\left(\mathbf{k}\right)\right|_{A,B,C,D=0,1,0,0}+i\left.\hat{\Gamma}_{B_{1}}\left(\mathbf{k}\right)\right|_{A,B,C,D=0,0,1,0}\label{eq:adiab_evol_3A1b}\\
\hat{\Gamma}_{^{3}A_{2}(b)}\left(\mathbf{k}\right) & = & \left.\hat{\Gamma}_{B_{1}}\left(\mathbf{k}\right)\right|_{A,B,C,D=0,1,0,0}+i\left.\hat{\Gamma}_{B_{2}}\left(\mathbf{k}\right)\right|_{A,B,C,D=0,0,1,0}\label{eq:adiab_evol_3A2b}\\
\hat{\Gamma}_{^{3}B_{1}(b)}\left(\mathbf{k}\right) & = & \left.\hat{\Gamma}_{A_{2}}\left(\mathbf{k}\right)\right|_{A,B,C,D=0,1,0,0}+i\left.\hat{\Gamma}_{A_{1}}\left(\mathbf{k}\right)\right|_{A,B,C,D=0,0,1,0}\label{eq:adiab_evol_3B1b}\\
\hat{\Gamma}_{^{3}B_{2}(b)}\left(\mathbf{k}\right) & = & \left.\hat{\Gamma}_{A_{1}}\left(\mathbf{k}\right)\right|_{A,B,C,D=0,1,0,0}+i\left.\hat{\Gamma}_{A_{2}}\left(\mathbf{k}\right)\right|_{A,B,C,D=0,0,1,0}\label{eq:adiab_evol_3B2b}\end{eqnarray}

\end{widetext}This implies that, unlike the singlet and unitary triplet
instabilities \emph{the nonunitary triplet instabilities split under
the influence of SOC}: as SOC is increased the critical temperature
$T_{c}$ splits into two transitions, one in which the order parameter
takes one form and a second one where another component develops.
The first transition does not break TRS, but the second one does (it
wouldn't if the system went into that state straight from the normal
state; TRS breaking is due to the presence of the other component
of the order parameter and their relative phase, which is fixed by
the requirement that the correct form is recovered in the limit of zero
SOC). In the limit of weak SOC, the two transitions happen so close
that they are indistinguishable from a single transition going straight
into the state with broken TRS. 

\section{Discussion\label{sec:Discussion}}

\begin{figure*}
\includegraphics[width=1.8\columnwidth,keepaspectratio]{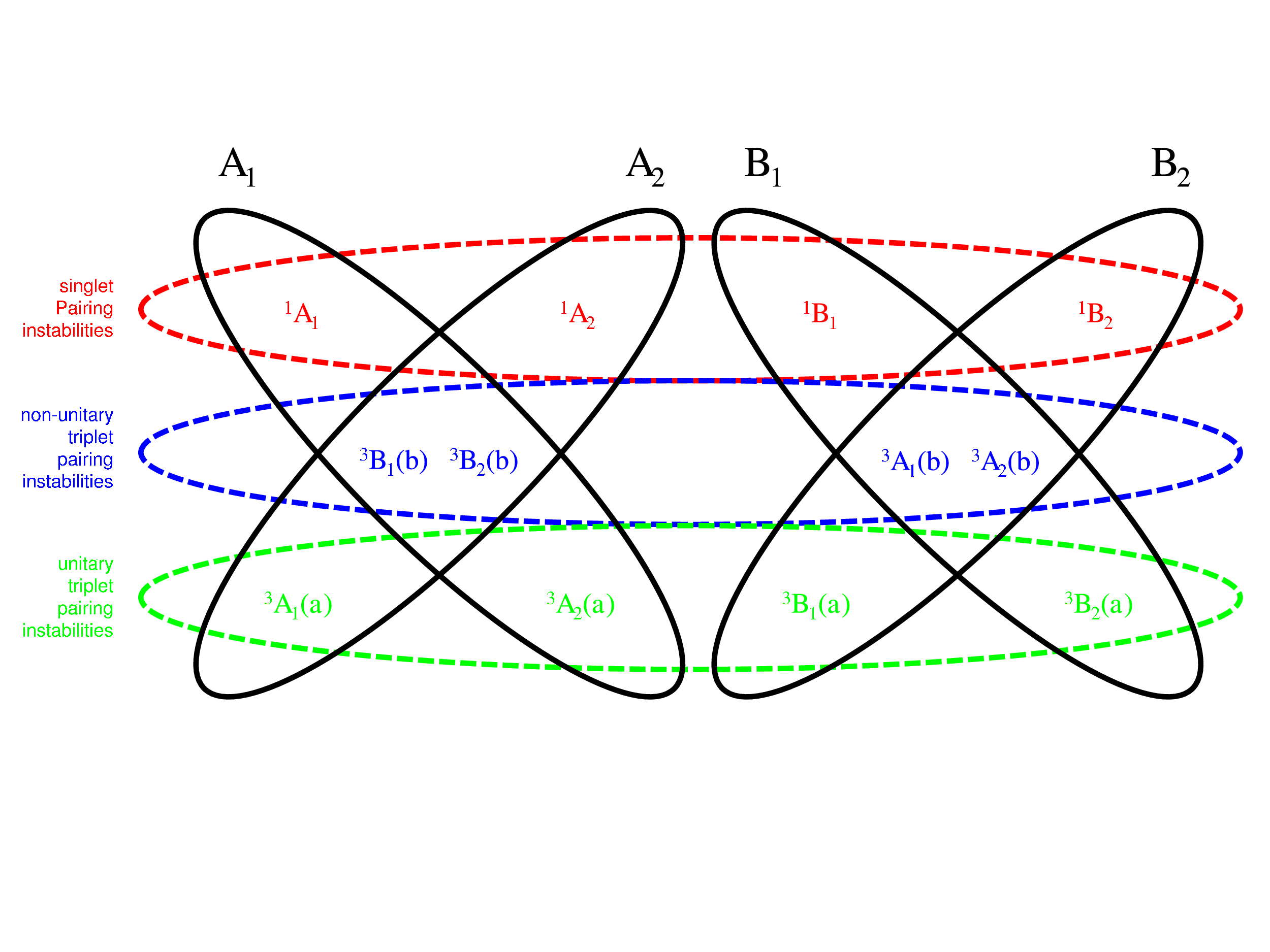}
\caption{\label{fig:Venn}(color online) Venn-Euler diagram showing how the four pairing states allowed just below $T_c$ in the presence of spin-orbit coupling decompose into the 12 singlet, non-unitary triplet and unitary triplet states states allowed in its absence.}
\end{figure*}

Figure \ref{fig:Evolution} shows the evolution of the superconducting instabilities allowed by symmetry in the absence of SOC as the latter is adiabtically turned on. We can pose the opposite question, which is: in the presence of strong SOC, how is a general pairing state decomposed into the components that would be allowed in its absence? This is shown in Fig. \ref{fig:Venn}. We note that in general the pairing states allowed in the presence of SOC contain singlet, unitary and non-untiary triplet components. Interestingly, the non-unitary states, which are the only ones that can break TRS, are always shared between two different strong SOC pairing states. On the other hand the singlet $s$-wave state never contributes to a TRS breaking instability. Also interestingly, as shown in Fig. \ref{fig:Venn}, the singlet $^1A_1$ state does mix with several triplet states, including part of the non-unitary triplet pairings $^3B_1(b)$ and $^3B_2(b)$. Nevertheless, and somewhat counter-intuitively, none of these combinations break TRS. 

In the light of the above analysis  let us now consider possible pairing states in LaNiC$_2$. The authors of Ref.~\onlinecite{subedi_electron-phonon_2009} have argued
that the normal state of LaNiC$_{2}$ is weakly correlated and that the
superconducting instability is of the conventional, $s$-wave type,
resulting from phonon-mediated pairing of electrons. The justification
provided for these assumptions is that a value of $T_{c}$ very close
to that encountered in the experiments follows from them. To explain
the observation \cite{hillier_evidence_2009} of TRS breaking, a small
triplet component induced by SOC is invoked. Indeed an order parameter
with $^{1}A_{1}$ symmetry would develop a small triplet component
as a result of SOC, as shown in Eq.~(\ref{eq:adiab_evol_1A1}) and Fig.~\ref{fig:Venn}. Unfortunately,
however, such triplet component is not TRS-breaking: only the non-unitary
triplet pairing instabilities can break TRS. These instabilities,
on the other hand, only have a small $s$-wave component, which vanishes
completely as SOC is turned off. 
%Thus the arguments presented in Ref.~\onlinecite{subedi_electron-phonon_2009}
%cannot be correct as they are based on a nonrelativsitic calculation
%and assume the vanishing in that limit of the triplet component. Indeed
Our results imply that \emph{only nonunitary triplet pairing is compatible
with the observation of TRS breaking.} 

A second consequence of our results, as shown in Fig.~\ref{fig:Evolution}, is that \emph{the superconducting
instability must be split by SOC.} Since this only happens for the
nonunitary triplet pairing instabilities, the observation of a split
transition would be a direct consequence of TRS breaking and confirm
the nonunitary triplet pairing in this system. On the other hand, given that it has not been detected
in any experiment to date, the splitting must be quite small. Its
observation may require the availability of single crystals, where
any splitting may be more easily observed.

An outstanding issue is the quantitative estimation of the size of SOC in LaNiC$_{2}$.
The band splitting has been calculated perturbatively using as the
starting point a band structure obtained in the local density approximation
(LDA).\cite{hase_electronic_2009} An average band splitting of $\sim3.1\mbox{mRy},$ about half the value of that obtained by a similar method in the noncentrosymmetric heavy fermion superconductor CePt$_{3}$Si,\cite{samokhin_cept3si:unconventional_2004,samokhin_erratum:_2004}
was found. Given that the critical temperature of LaNiC$_{2}$ is
about three times higher than that of CePt$_{3}$Si this suggests
that the possible role played by SOC in LaNiC$_{2}$ is smaller.
That said, even in this case the obtained splitting is an average, and for some parts of the Fermi surface it can be either larger or smaller than that value.\cite{hase_electronic_2009} The importance of SOC thus depends on a number of details that are as yet unkwon, such as the exact functional form of the superconducting order parameter. In any case the average value is much larger
than the superconducting gap, and of the same order of magnitude as
the Debye energy.\cite{hase_electronic_2009} Yet as we have shown
above if SOC had a strong effect on the superconducting instability the latter would
not break TRS, which is at variance with the experimental data.\cite{hillier_evidence_2009}
%This, together with the unconventional symmetry of the order parameter
%and the fact that the density of states near the Fermi energy is dominated
%by Ni 3$d$ orbitals, suggests that the normal state LaNiC$_{2}$
%is strongly correlated and not adequately described in the LDA. 
We note that LDA-based estimates of SOC have been called into question in the case of the heavy fermion noncentrosymmetric superconductor CePt$_{3}$Si\cite{hasegawa_spin_2009}
where
de Haas-van Alphen oscillations have failed to detect the predicted
band splitting.\cite{2004-Hashimoto}

%\textbf{{[}}\textbf{\underbar{Adrian,}} \textbf{could you like to
%include here the discussion of the Pauli limit? We could have a discussion
%similar to that in Ref.~}\cite{powell_phenomenological_2008} \textbf{with
%the values of $H_{c2}$ and $H_{Pauli}$ that you obtained. However
%I am rather in favour of keeping that for the paper on trasnverse
%$\mu$SR.]} 

All discussions so far of the implications of the observation
of TRS breaking in LaNiC$_{2}$, \cite{hillier_evidence_2009,subedi_electron-phonon_2009}
including the one presented here, assume that this is a bulk phenomenon.
However albeit very pure, the samples on which this was observed were
polycrystalline.\cite{hillier_evidence_2009} A distinct possibility
is that the observations could correspond to a breaking of TRS at
the boundaries between crystallites.\JQ{\cite{mineev_private}} On such surfaces the crystal symmetry is broken and the list of
symmetry-allowed superconducting instabilities is altered. On the
other hand in the experiment described in Ref.~\onlinecite{hillier_evidence_2009}
muons were deposited uniformly throughout the bulk of the sample.
Any magnetic fields occurring only at the boundaries between crystallites
would have been screened over distances of the order of the penetration
depth, $\lambda$. In order to discard completely this possibility it would therefore be required to know this number, which can be obtained for example in a transverse-field $\mu$SR experiment.

\section{Conclusion}

In conclusion, we have studied, on the basis of group-theoretical
considerations, the effect of spin-orbit coupling (SOC) of arbitrary
strength on the superconducting instability of the noncentrosymmetric
intermetallic compound LaNiC$_{2}$. We have paid particular attention
to the issue of time reversal symmetry (TRS) breaking. While in the
absence of SOC there are 12 possible superconducting instabilities,
of which 4 break TRS, when SOC is taken into account there are only
4 superconducting instabilities of the normal state, and none of them
break TRS. To reconcile this result with the experimental observation
of TRS breaking on entering the superconducting state \cite{hillier_evidence_2009}
we have studied the evolution of the superconducting instability as
a small amount of SOC is adiabatically turned on. We have found that
each of the 8 TRS-preserving singlet and unitary triplet instabilities
evolve smoothly into one of the 4 that are allowed in the presence
of SOC and we have obtained the form these must take when SOC is small,
but finite. In particular our analysis shows a small triplet component
developing on top of an $s$-wave order parameter. However, this mechanism
is found \emph{not} to lead to TRS breaking. A similar analysis for
the case of the 4 nonunitary triplet pairing instabilities reveals
that each of them splits into two distinct transitions: an instability
of the normal state where superconductivity emerges without the breaking
of TRS, followed by a second superconducting instability where the
order parameter acquires an additional component and TRS is broken.
We thus conclude that the superconducting instability must be of the
nonunitary triplet type and that SOC must be comparatively small in
this system so as to make the first and second transitions indiscernible.
A distinct prediction of this analysis is a splitting of the superconducting
transition that could be observed in single crystals and enhanced
by the application of pressure. Since only the nonunitary triplet
pairing instabilities are split in this way, its observation would
be a direct consequence of the broken TRS and constitute definitive
proof of nonunitary triplet pairing in this system.

Non-unitary triplet pairing is believed to be realised in the ferromagnetic superconductors.\cite{UGe2,URhGe,UIr,UCoGe,de_Visser_review} In contrast, the normal state of LaNiC$_{2}$ just above $T_c$ is paramagnetic. This material therefore constitutes the first example of an ``intrinsically'' nonunitary
triplet superconductor where the pairing of electrons with only one value of the spin does not result from a pre-existing exchange splitting. Elucidating the mechanism by which this comes about, and the possible role that the lack of inversion symmetry may play in it, is an outstanding challenge. 

\acknowledgments

The authors thank Vladimir P. Mineev for a stimulating discussion. JQ gratefully acknowledges funding from STFC in association with St. Catherine's College, Oxford and from HEFCE through the South-East Physics network (SEPnet).


\begin{thebibliography}{38}
\expandafter\ifx\csname natexlab\endcsname\relax\def\natexlab#1{#1}\fi
\expandafter\ifx\csname bibnamefont\endcsname\relax
  \def\bibnamefont#1{#1}\fi
\expandafter\ifx\csname bibfnamefont\endcsname\relax
  \def\bibfnamefont#1{#1}\fi
\expandafter\ifx\csname citenamefont\endcsname\relax
  \def\citenamefont#1{#1}\fi
\expandafter\ifx\csname url\endcsname\relax
  \def\url#1{\texttt{#1}}\fi
\expandafter\ifx\csname urlprefix\endcsname\relax\def\urlprefix{URL }\fi
\providecommand{\bibinfo}[2]{#2}
\providecommand{\eprint}[2][]{\url{#2}}

\bibitem[{\citenamefont{Bauer et~al.}(2004)}]{bauer_heavy_2004}
\bibinfo{author}{\bibfnamefont{E.}~\bibnamefont{Bauer}} \bibnamefont{et~al.},
  \bibinfo{journal}{Physical Review Letters} \textbf{\bibinfo{volume}{92}},
  \bibinfo{pages}{027003} (\bibinfo{year}{2004}).

\bibitem[{\citenamefont{Anderson}(1984)}]{anderson_1984}
\bibinfo{author}{\bibfnamefont{P.~W.} \bibnamefont{Anderson}},
  \bibinfo{journal}{Phys. Rev. B} \textbf{\bibinfo{volume}{30}}
  (\bibinfo{year}{1984}).

\bibitem[{\citenamefont{Volovik and Gor'kov}(1984)}]{volovik_1984}
\bibinfo{author}{\bibfnamefont{G.~E.} \bibnamefont{Volovik}} \bibnamefont{and}
  \bibinfo{author}{\bibfnamefont{L.~P.} \bibnamefont{Gor'kov}},
  \bibinfo{journal}{Pis'ma Zh. Eksp. Teor. Fiz.} \textbf{\bibinfo{volume}{39}}
  (\bibinfo{year}{1984}), \bibinfo{note}{[English version: JETP Letters {\bf
  39}, 674 (1984)]}.

\bibitem[{\citenamefont{Ueda and Rice}(1985)}]{ueda_rice_1985}
\bibinfo{author}{\bibfnamefont{K.}~\bibnamefont{Ueda}} \bibnamefont{and}
  \bibinfo{author}{\bibfnamefont{T.~M.} \bibnamefont{Rice}},
  \bibinfo{journal}{Phys. Rev. B} \textbf{\bibinfo{volume}{31}}
  (\bibinfo{year}{1985}).

\bibitem[{\citenamefont{aki Ozaki et~al.}(1985)\citenamefont{aki Ozaki,
  Machida, and Ohmi}}]{ozaki_machida_omi_1985}
\bibinfo{author}{\bibfnamefont{M.}~\bibnamefont{aki Ozaki}},
  \bibinfo{author}{\bibfnamefont{K.}~\bibnamefont{Machida}}, \bibnamefont{and}
  \bibinfo{author}{\bibfnamefont{T.}~\bibnamefont{Ohmi}},
  \bibinfo{journal}{Prog. Theor. Phys.} \textbf{\bibinfo{volume}{74}}
  (\bibinfo{year}{1985}).

\bibitem[{\citenamefont{Blount}(1985)}]{blount_1985}
\bibinfo{author}{\bibfnamefont{E.~I.} \bibnamefont{Blount}},
  \bibinfo{journal}{Phys. Rev. B} \textbf{\bibinfo{volume}{32}}
  (\bibinfo{year}{1985}).

\bibitem[{\citenamefont{Weinberg}(1995)}]{Weinberg}
\bibinfo{author}{\bibfnamefont{S.}~\bibnamefont{Weinberg}},
  \emph{\bibinfo{title}{The Quantum Theory of Fields}}
  (\bibinfo{publisher}{Cambridge University Press}, \bibinfo{year}{1995}).

\bibitem[{\citenamefont{Yuan et~al.}(2006)}]{Yuan_Li2Pd3B_2006}
\bibinfo{author}{\bibfnamefont{H.~Q.} \bibnamefont{Yuan}} \bibnamefont{et~al.},
  \bibinfo{journal}{Phys. Rev. Lett.} \textbf{\bibinfo{volume}{97}}
  (\bibinfo{year}{2006}).

\bibitem[{\citenamefont{Bauer et~al.}(2009)}]{Bauer_BaPtSi3_2009}
\bibinfo{author}{\bibfnamefont{E.}~\bibnamefont{Bauer}} \bibnamefont{et~al.},
  \bibinfo{journal}{Phys. Rev. B} \textbf{\bibinfo{volume}{80}}
  (\bibinfo{year}{2009}).

\bibitem[{\citenamefont{Zuev et~al.}(2007)\citenamefont{Zuev, Kuznetsova,
  Prozorov, Vannette, Lobanov, Christen, and Thompson}}]{Zuev_Re3W_2007}
\bibinfo{author}{\bibfnamefont{Y.~L.} \bibnamefont{Zuev}},
  \bibinfo{author}{\bibfnamefont{V.~A.} \bibnamefont{Kuznetsova}},
  \bibinfo{author}{\bibfnamefont{R.}~\bibnamefont{Prozorov}},
  \bibinfo{author}{\bibfnamefont{M.~D.} \bibnamefont{Vannette}},
  \bibinfo{author}{\bibfnamefont{M.~V.} \bibnamefont{Lobanov}},
  \bibinfo{author}{\bibfnamefont{D.~K.} \bibnamefont{Christen}},
  \bibnamefont{and} \bibinfo{author}{\bibfnamefont{J.~R.}
  \bibnamefont{Thompson}}, \bibinfo{journal}{Phys. Rev. B}
  \textbf{\bibinfo{volume}{76}}, \bibinfo{pages}{132508}
  (\bibinfo{year}{2007}).

\bibitem[{\citenamefont{Hillier et~al.}(2009)\citenamefont{Hillier,
  Quintanilla, and Cywinski}}]{hillier_evidence_2009}
\bibinfo{author}{\bibfnamefont{A.~D.} \bibnamefont{Hillier}},
  \bibinfo{author}{\bibfnamefont{J.}~\bibnamefont{Quintanilla}},
  \bibnamefont{and} \bibinfo{author}{\bibfnamefont{R.}~\bibnamefont{Cywinski}},
  \bibinfo{journal}{Physical Review Letters} \textbf{\bibinfo{volume}{102}},
  \bibinfo{pages}{117007} (\bibinfo{year}{2009}).

\bibitem[{\citenamefont{Bodak and Marusin}(1979)}]{bodak_marusin}
\bibinfo{author}{\bibfnamefont{O.~I.} \bibnamefont{Bodak}} \bibnamefont{and}
  \bibinfo{author}{\bibfnamefont{E.~P.} \bibnamefont{Marusin}},
  \bibinfo{journal}{Dopovidi-Akademii Nauk Ukrains'koi RSR Seriia A}
  (\bibinfo{year}{1979}).

\bibitem[{\citenamefont{Lee et~al.}(1996)\citenamefont{Lee, Zeng, Yao, and
  Chen}}]{lee_superconductivity_1996}
\bibinfo{author}{\bibfnamefont{W.~H.} \bibnamefont{Lee}},
  \bibinfo{author}{\bibfnamefont{H.~K.} \bibnamefont{Zeng}},
  \bibinfo{author}{\bibfnamefont{Y.~D.} \bibnamefont{Yao}}, \bibnamefont{and}
  \bibinfo{author}{\bibfnamefont{Y.~Y.} \bibnamefont{Chen}},
  \bibinfo{journal}{Physica C: Superconductivity}
  \textbf{\bibinfo{volume}{266}}, \bibinfo{pages}{138} (\bibinfo{year}{1996}),
  ISSN \bibinfo{issn}{0921-4534}.

\bibitem[{\citenamefont{Pecharsky et~al.}(1998)\citenamefont{Pecharsky, Miller,
  and Gschneidner}}]{pecharsky_low-temperature_1998}
\bibinfo{author}{\bibfnamefont{V.~K.} \bibnamefont{Pecharsky}},
  \bibinfo{author}{\bibfnamefont{L.~L.} \bibnamefont{Miller}},
  \bibnamefont{and} \bibinfo{author}{\bibfnamefont{K.~A.}
  \bibnamefont{Gschneidner}}, \bibinfo{journal}{Physical Review B}
  \textbf{\bibinfo{volume}{58}}, \bibinfo{pages}{497} (\bibinfo{year}{1998}).

\bibitem[{\citenamefont{Sung et~al.}(2008)\citenamefont{Sung, Chou, Syu, and
  Lee}}]{sung_antipressure_2008}
\bibinfo{author}{\bibfnamefont{H.~H.} \bibnamefont{Sung}},
  \bibinfo{author}{\bibfnamefont{S.~Y.} \bibnamefont{Chou}},
  \bibinfo{author}{\bibfnamefont{K.~J.} \bibnamefont{Syu}}, \bibnamefont{and}
  \bibinfo{author}{\bibfnamefont{W.~H.} \bibnamefont{Lee}},
  \bibinfo{journal}{Journal of Physics: Condensed Matter}
  \textbf{\bibinfo{volume}{20}}, \bibinfo{pages}{165207}
  (\bibinfo{year}{2008}), ISSN \bibinfo{issn}{0953-8984}.

\bibitem[{\citenamefont{Liao et~al.}(2009)\citenamefont{Liao, Sung, Syu, and
  Lee}}]{t.f._liao_alloying_2009}
\bibinfo{author}{\bibfnamefont{T.}~\bibnamefont{Liao}},
  \bibinfo{author}{\bibfnamefont{H.~H.} \bibnamefont{Sung}},
  \bibinfo{author}{\bibfnamefont{K.~J.} \bibnamefont{Syu}}, \bibnamefont{and}
  \bibinfo{author}{\bibfnamefont{W.~H.} \bibnamefont{Lee}},
  \bibinfo{journal}{Solid State Communications} \textbf{\bibinfo{volume}{149}},
  \bibinfo{pages}{448} (\bibinfo{year}{2009}).

\bibitem[{\citenamefont{Hase and Yanagisawa}(2009)}]{hase_electronic_2009}
\bibinfo{author}{\bibfnamefont{I.}~\bibnamefont{Hase}} \bibnamefont{and}
  \bibinfo{author}{\bibfnamefont{T.}~\bibnamefont{Yanagisawa}},
  \bibinfo{journal}{Journal of the Physical Society of Japan}
  \textbf{\bibinfo{volume}{78}}, \bibinfo{pages}{084724}
  (\bibinfo{year}{2009}), ISSN \bibinfo{issn}{0031-9015}.

\bibitem[{\citenamefont{Laverock et~al.}(2009)\citenamefont{Laverock, Haynes,
  Utfeld, and Dugdale}}]{laverock_electronic_2009}
\bibinfo{author}{\bibfnamefont{J.}~\bibnamefont{Laverock}},
  \bibinfo{author}{\bibfnamefont{T.~D.} \bibnamefont{Haynes}},
  \bibinfo{author}{\bibfnamefont{C.}~\bibnamefont{Utfeld}}, \bibnamefont{and}
  \bibinfo{author}{\bibfnamefont{S.~B.} \bibnamefont{Dugdale}},
  \bibinfo{journal}{Physical Review B {(Condensed} Matter and Materials
  Physics)} \textbf{\bibinfo{volume}{80}}, \bibinfo{pages}{125111}
  (\bibinfo{year}{2009}).

\bibitem[{\citenamefont{Subedi and Singh}(2009)}]{subedi_electron-phonon_2009}
\bibinfo{author}{\bibfnamefont{A.}~\bibnamefont{Subedi}} \bibnamefont{and}
  \bibinfo{author}{\bibfnamefont{D.~J.} \bibnamefont{Singh}},
  \bibinfo{journal}{Physical Review B {(Condensed} Matter and Materials
  Physics)} \textbf{\bibinfo{volume}{80}}, \bibinfo{pages}{092506}
  (\bibinfo{year}{2009}).

\bibitem[{\citenamefont{Lee and Zeng}(1997)}]{Lee1997323}
\bibinfo{author}{\bibfnamefont{W.~H.} \bibnamefont{Lee}} \bibnamefont{and}
  \bibinfo{author}{\bibfnamefont{H.~K.} \bibnamefont{Zeng}},
  \bibinfo{journal}{Solid State Communications} \textbf{\bibinfo{volume}{101}},
  \bibinfo{pages}{323 } (\bibinfo{year}{1997}).

\bibitem[{\citenamefont{Lee et~al.}(1997)\citenamefont{Lee, Zeng, Chen, Yao,
  and Ho}}]{Lee1997433}
\bibinfo{author}{\bibfnamefont{W.~H.} \bibnamefont{Lee}},
  \bibinfo{author}{\bibfnamefont{H.~K.} \bibnamefont{Zeng}},
  \bibinfo{author}{\bibfnamefont{Y.~Y.} \bibnamefont{Chen}},
  \bibinfo{author}{\bibfnamefont{Y.~D.} \bibnamefont{Yao}}, \bibnamefont{and}
  \bibinfo{author}{\bibfnamefont{J.~C.} \bibnamefont{Ho}},
  \bibinfo{journal}{Solid State Communications} \textbf{\bibinfo{volume}{102}},
  \bibinfo{pages}{433 } (\bibinfo{year}{1997}).

\bibitem[{\citenamefont{Mineev and Samokhin}(1999)}]{mineev_introduction_1999}
\bibinfo{author}{\bibfnamefont{V.~P.} \bibnamefont{Mineev}} \bibnamefont{and}
  \bibinfo{author}{\bibfnamefont{K.~V.} \bibnamefont{Samokhin}},
  \emph{\bibinfo{title}{Introduction to Unconventional Superconductivity}}
  (\bibinfo{publisher}{Taylor \& Francis Ltd}, \bibinfo{year}{1999}).

\bibitem[{\citenamefont{Annett}(1990)}]{james_f_annett_symmetry_1990}
\bibinfo{author}{\bibfnamefont{J.~F.} \bibnamefont{Annett}},
  \bibinfo{journal}{{ADVANCES} {IN} {PHYSICS}} \textbf{\bibinfo{volume}{39}},
  \bibinfo{pages}{83} (\bibinfo{year}{1990}).

\bibitem[{\citenamefont{Sergienko and Curnoe}(2004)}]{2004-Sergienko}
\bibinfo{author}{\bibfnamefont{I.~A.} \bibnamefont{Sergienko}}
  \bibnamefont{and} \bibinfo{author}{\bibfnamefont{S.~H.}
  \bibnamefont{Curnoe}}, \bibinfo{journal}{Physical Review B}
  \textbf{\bibinfo{volume}{70}}, \bibinfo{pages}{214510}
  (\bibinfo{year}{2004}), \bibinfo{note}{copyright {(C)} 2009 The American
  Physical Society; Please report any problems to prola@aps.org},
  \urlprefix\url{http://link.aps.org/abstract/PRB/v70/e214510}.

\bibitem[{\citenamefont{Samokhin
  et~al.}(2004{\natexlab{a}})\citenamefont{Samokhin, Zijlstra, and
  Bose}}]{samokhin_cept3si:unconventional_2004}
\bibinfo{author}{\bibfnamefont{K.~V.} \bibnamefont{Samokhin}},
  \bibinfo{author}{\bibfnamefont{E.~S.} \bibnamefont{Zijlstra}},
  \bibnamefont{and} \bibinfo{author}{\bibfnamefont{S.~K.} \bibnamefont{Bose}},
  \bibinfo{journal}{Physical Review B} \textbf{\bibinfo{volume}{69}},
  \bibinfo{pages}{094514} (\bibinfo{year}{2004}{\natexlab{a}}).

\bibitem[{\citenamefont{Samokhin
  et~al.}(2004{\natexlab{b}})\citenamefont{Samokhin, Zijlstra, and
  Bose}}]{samokhin_erratum:_2004}
\bibinfo{author}{\bibfnamefont{K.~V.} \bibnamefont{Samokhin}},
  \bibinfo{author}{\bibfnamefont{E.~S.} \bibnamefont{Zijlstra}},
  \bibnamefont{and} \bibinfo{author}{\bibfnamefont{S.~K.} \bibnamefont{Bose}},
  \bibinfo{journal}{Physical Review B} \textbf{\bibinfo{volume}{70}},
  \bibinfo{pages}{069902} (\bibinfo{year}{2004}{\natexlab{b}}).

\bibitem[{\citenamefont{Sigrist and
  Ueda}(1991)}]{sigrist_phenomenological_1991}
\bibinfo{author}{\bibfnamefont{M.}~\bibnamefont{Sigrist}} \bibnamefont{and}
  \bibinfo{author}{\bibfnamefont{K.}~\bibnamefont{Ueda}},
  \bibinfo{journal}{Reviews of Modern Physics} \textbf{\bibinfo{volume}{63}},
  \bibinfo{pages}{239} (\bibinfo{year}{1991}).

\bibitem[{\citenamefont{Gor'kov and
  Rashba}(2001)}]{gorkov_superconducting_2001}
\bibinfo{author}{\bibfnamefont{L.~P.} \bibnamefont{Gor'kov}} \bibnamefont{and}
  \bibinfo{author}{\bibfnamefont{E.~I.} \bibnamefont{Rashba}},
  \bibinfo{journal}{Physical Review Letters} \textbf{\bibinfo{volume}{87}},
  \bibinfo{pages}{037004} (\bibinfo{year}{2001}).

\bibitem[{\citenamefont{Frigeri
  et~al.}(2004{\natexlab{a}})\citenamefont{Frigeri, Agterberg, Koga, and
  Sigrist}}]{frigeri_erratum:_2004}
\bibinfo{author}{\bibfnamefont{P.~A.} \bibnamefont{Frigeri}},
  \bibinfo{author}{\bibfnamefont{D.~F.} \bibnamefont{Agterberg}},
  \bibinfo{author}{\bibfnamefont{A.}~\bibnamefont{Koga}}, \bibnamefont{and}
  \bibinfo{author}{\bibfnamefont{M.}~\bibnamefont{Sigrist}},
  \bibinfo{journal}{Physical Review Letters} \textbf{\bibinfo{volume}{93}},
  \bibinfo{pages}{099903} (\bibinfo{year}{2004}{\natexlab{a}}).

\bibitem[{\citenamefont{Frigeri
  et~al.}(2004{\natexlab{b}})\citenamefont{Frigeri, Agterberg, Koga, and
  Sigrist}}]{frigeri_superconductivity_2004}
\bibinfo{author}{\bibfnamefont{P.~A.} \bibnamefont{Frigeri}},
  \bibinfo{author}{\bibfnamefont{D.~F.} \bibnamefont{Agterberg}},
  \bibinfo{author}{\bibfnamefont{A.}~\bibnamefont{Koga}}, \bibnamefont{and}
  \bibinfo{author}{\bibfnamefont{M.}~\bibnamefont{Sigrist}},
  \bibinfo{journal}{Physical Review Letters} \textbf{\bibinfo{volume}{92}},
  \bibinfo{pages}{097001} (\bibinfo{year}{2004}{\natexlab{b}}).

\bibitem[{\citenamefont{Hasegawa and Taniguchi}(2009)}]{hasegawa_spin_2009}
\bibinfo{author}{\bibfnamefont{Y.}~\bibnamefont{Hasegawa}} \bibnamefont{and}
  \bibinfo{author}{\bibfnamefont{H.}~\bibnamefont{Taniguchi}},
  \bibinfo{journal}{Journal of the Physical Society of Japan}
  \textbf{\bibinfo{volume}{78}}, \bibinfo{pages}{074717}
  (\bibinfo{year}{2009}), ISSN \bibinfo{issn}{0031-9015}.

\bibitem[{\citenamefont{Hashimoto et~al.}(2004)\citenamefont{Hashimoto, Yasuda,
  Kubo, Shishido, Ueda, Settai, Matsuda, Haga, Harima, and
  Onuki}}]{2004-Hashimoto}
\bibinfo{author}{\bibfnamefont{S.}~\bibnamefont{Hashimoto}},
  \bibinfo{author}{\bibfnamefont{T.}~\bibnamefont{Yasuda}},
  \bibinfo{author}{\bibfnamefont{T.}~\bibnamefont{Kubo}},
  \bibinfo{author}{\bibfnamefont{H.}~\bibnamefont{Shishido}},
  \bibinfo{author}{\bibfnamefont{T.}~\bibnamefont{Ueda}},
  \bibinfo{author}{\bibfnamefont{R.}~\bibnamefont{Settai}},
  \bibinfo{author}{\bibfnamefont{T.~D.} \bibnamefont{Matsuda}},
  \bibinfo{author}{\bibfnamefont{Y.}~\bibnamefont{Haga}},
  \bibinfo{author}{\bibfnamefont{H.}~\bibnamefont{Harima}}, \bibnamefont{and}
  \bibinfo{author}{\bibfnamefont{Y.}~\bibnamefont{Onuki}},
  \bibinfo{journal}{Journal of Physics: Condensed Matter}
  \textbf{\bibinfo{volume}{16}}, \bibinfo{pages}{L287} (\bibinfo{year}{2004}),
  ISSN \bibinfo{issn}{0953-8984},
  \urlprefix\url{http://www.iop.org/EJ/abstract/0953-8984/16/23/L02/}.

\bibitem[{\citenamefont{Mineev}(2009)}]{mineev_private}
\bibinfo{author}{\bibfnamefont{V.~P.} \bibnamefont{Mineev}},
  \bibinfo{howpublished}{private communication} (\bibinfo{year}{2009}).

\bibitem[{\citenamefont{Pfleiderer and Huxley}(2002)}]{UGe2}
\bibinfo{author}{\bibfnamefont{C.}~\bibnamefont{Pfleiderer}} \bibnamefont{and}
  \bibinfo{author}{\bibfnamefont{A.~D.} \bibnamefont{Huxley}},
  \bibinfo{journal}{Phys. Rev. Lett.} \textbf{\bibinfo{volume}{89}}
  (\bibinfo{year}{2002}).

\bibitem[{\citenamefont{Hardy et~al.}(2005)\citenamefont{Hardy, Huxley,
  Flouquet, Salce, Knebel, Braithwaite, Aoki, Uhlarz, and Pfleiderer}}]{URhGe}
\bibinfo{author}{\bibfnamefont{F.}~\bibnamefont{Hardy}},
  \bibinfo{author}{\bibfnamefont{A.~D.} \bibnamefont{Huxley}},
  \bibinfo{author}{\bibfnamefont{J.}~\bibnamefont{Flouquet}},
  \bibinfo{author}{\bibfnamefont{B.}~\bibnamefont{Salce}},
  \bibinfo{author}{\bibfnamefont{G.}~\bibnamefont{Knebel}},
  \bibinfo{author}{\bibfnamefont{D.}~\bibnamefont{Braithwaite}},
  \bibinfo{author}{\bibfnamefont{D.}~\bibnamefont{Aoki}},
  \bibinfo{author}{\bibfnamefont{M.}~\bibnamefont{Uhlarz}}, \bibnamefont{and}
  \bibinfo{author}{\bibfnamefont{C.}~\bibnamefont{Pfleiderer}},
  \bibinfo{journal}{Physica B} \textbf{\bibinfo{volume}{359-61}}
  (\bibinfo{year}{2005}).

\bibitem[{\citenamefont{Kobayashi et~al.}(2006)\citenamefont{Kobayashi,
  Fukushima, Hidaka, Kotegawa, Akazawa, Yamamoto, Haga, Settai, and
  Onuki}}]{UIr}
\bibinfo{author}{\bibfnamefont{T.~C.} \bibnamefont{Kobayashi}},
  \bibinfo{author}{\bibfnamefont{S.}~\bibnamefont{Fukushima}},
  \bibinfo{author}{\bibfnamefont{H.}~\bibnamefont{Hidaka}},
  \bibinfo{author}{\bibfnamefont{H.}~\bibnamefont{Kotegawa}},
  \bibinfo{author}{\bibfnamefont{T.}~\bibnamefont{Akazawa}},
  \bibinfo{author}{\bibfnamefont{E.}~\bibnamefont{Yamamoto}},
  \bibinfo{author}{\bibfnamefont{Y.}~\bibnamefont{Haga}},
  \bibinfo{author}{\bibfnamefont{R.}~\bibnamefont{Settai}}, \bibnamefont{and}
  \bibinfo{author}{\bibfnamefont{Y.}~\bibnamefont{Onuki}},
  \bibinfo{journal}{Physica B} \textbf{\bibinfo{volume}{378--80}}
  (\bibinfo{year}{2006}).

\bibitem[{\citenamefont{Huy et~al.}(2008)\citenamefont{Huy, de~Nijs, Huang, and
  de~Visser}}]{UCoGe}
\bibinfo{author}{\bibfnamefont{N.~T.} \bibnamefont{Huy}},
  \bibinfo{author}{\bibfnamefont{D.~E.} \bibnamefont{de~Nijs}},
  \bibinfo{author}{\bibfnamefont{Y.}~\bibnamefont{Huang}}, \bibnamefont{and}
  \bibinfo{author}{\bibfnamefont{A.}~\bibnamefont{de~Visser}},
  \bibinfo{journal}{Phys. Rev. Lett.} \textbf{\bibinfo{volume}{100}}
  (\bibinfo{year}{2008}).

\bibitem[{\citenamefont{de~Visser}(2010)}]{de_Visser_review}
\bibinfo{author}{\bibfnamefont{A.}~\bibnamefont{de~Visser}}, in
  \emph{\bibinfo{booktitle}{Encyclopedia of Materials: Science and
  Technology}}, edited by \bibinfo{editor}{\bibfnamefont{K.~H.~J.}
  \bibnamefont{Buschow}} \bibnamefont{et~al.} (\bibinfo{publisher}{Elsevier},
  \bibinfo{year}{2010}), pp. \bibinfo{pages}{1--6}.

\end{thebibliography}
\end{document}